\documentclass[aps,epsf,twocolumn,showpacs,preprintnumbers]{revtex4}
\usepackage{graphics}
\usepackage{graphicx}
\usepackage{dcolumn} 
\usepackage{bm}
\usepackage{epsfig}
\usepackage{float}
\newcommand{\degree}{\ensuremath{^\circ}}

\begin{document}
\title{Quasi one-dimensional magnetism driven by unusual orbital ordering in 
CuSb$_{2}$O$_{6}$}
\author{Deepa Kasinathan,$^1$ 
  Klaus Koepernik,$^{1,2}$ and Helge Rosner$^1$}
\affiliation{$^1$Max-Planck-Institut f\"ur Chemische Physik fester
Stoffe Dresden, Germany }
\affiliation{$^2$IFW Dresden, P.O. Box 270116, D-01171 Dresden, Germany}
\date{\today}
\begin{abstract}
Essentially all undoped cuprates exhibit a quasi-planar, fourfold Cu-O
coordination responsible for the magnetically active antibonding
$3d_{x^2-y^2}$-like state.  Here, we present an electronic structure study
for CuSb$_{2}$O$_{6}$ that reveals, in contrast, a half-filled
3d$_{3z^2-r^2}$ orbital. This hitherto unobserved ground state
originates from a competition of in- and out-of-plaquette orbitals where
the strong Coulomb repulsion drives the surprising and unique orbital
ordering. This, in turn, gives rise to an unexpected quasi
one-dimensional magnetic behavior. Our results provide a consistent
explanation of recent thermodynamical and neutron diffraction measurements.
\end{abstract}

\maketitle
Low dimensional systems have always been of fundamental interest to
both experimentalists and theorists for their peculiar magnetic
properties. In recent years, especially boosted by the discovery of
the high-$T_c$ superconductivity in cuprates, low dimensional magnets
related to this family of compounds have been widely studied. Low
dimensional cuprates exhibit a large variety of exotic physical
properties. This variety is a result of the complex interplay of different
interactions; mainly covalency, ligand-fields and strong correlation
effects. 

A nearly universal component of cuprate systems is a strongly
elongated CuO$_{6}$-octahedron\cite{buschbaum} wherein the exotic behavior finds its
origin in the deceivingly simple planar, half-filled Cu-O$_4$ orbital
lying in its basal plane. Here, Cu 3$d$ and O 2$p$ states form a well
separated, half filled $pd-\sigma$ molecular-like orbital of $x^2-y^2$
symmetry that is magnetically active. In general, the low-lying
magnetic excitations, especially the magnetic ground state can be well
described within this subspace of the whole Hilbert space. This way,
the low-dimensional magnetic properties of a compound depend crucially on the arrangements
of these Cu-O$_4$ units forming different networks, for instance
(i) quasi 1D chains from isolated ({\it i.e.} Sr$_{2}$Cu(PO$_{4}$)$_{2}$\cite{belik,johannes})
edge-sharing ({\it i.e.} Li$_{2}$CuO$_{2}$\cite{weht,rosner99})or corner-sharing 
({\it i.e.} Sr$_{2}$CuO$_{3}$\cite{ami,rosner97}) CuO$_{4}$ plaquettes,
 (ii) quasi 2D square lattices in La$_2$CuO$_4$\cite{shirane}, etc.


In contrast to the vast majority of cuprates with a quasi-planar
Cu(II) four-fold coordination, copper diantimonate CuSb$_{2}$O$_{6}$
-- first synthesized in early 1940's\cite{bystrom} -- exhibits a
nearly octahedral local environment of the Cu$^{2+}$ ions.  The
compound is green in color\cite{bystrom,nakua}, indicating insulating
behavior with a gap size typical for cuprates.  Although the compound
undergoes a second order phase transition below 380 K from a
tetragonal to a monoclinically distorted structure,\cite{giere} this
quasi-octahedral local environment is basically unchanged. The
magnetic cation sublattice is that of the K$_{2}$NiF$_{4}$ structure
type, which includes many classic examples of square lattices
exhibiting 2D antiferromagnetism (AFM), for instance the above
mentioned La$_2$CuO$_4$. Surprisingly, susceptibility measurements
done on both powder and single crystal samples of CuSb$_{2}$O$_{6}$
fit extremly well over a large temperature range to a
nearest-neighbor-only S = $\frac{1}{2}$ Heisenberg 1D model with an
exchange constant ranging from -86 K to -98
K.\cite{nakua,yamaguchi,kato,heinrich,prokofiev,gibson,footnote}
Furthermore, all low temperature susceptibility measurements show a
sharp drop at
8.5K\cite{nakua,yamaguchi,kato,heinrich,prokofiev,gibson}, due to the
onset of long range AFM ordering.\cite{heinrich,prokofiev,gibson}
Thus, the contrast of the 2D lattice and the 1D magnetic behavior in
CuSb$_{2}$O$_{6}$ has been puzzling up to now.

In this paper we
present the combined results of the total energy and tight-binding model (TBM) calculations,
which indicate an hitherto unobserved ground state originating from a 
competition between the in- and out-of-plaquette orbitals. 
Strong Coulomb correlation drives a surprising and
unique orbital order in CuSb$_{2}$O$_{6}$, thereby leading to the
strongly one-dimensional magnetic behavior with exchange integrals in 
good agreement with those deduced from experiments. Our results are
consistent with recent neutron diffraction measurements.\cite{gibson}

\begin{figure}[t]
\begin{center}\includegraphics[%
  clip,
  width=8cm,
  angle=0]{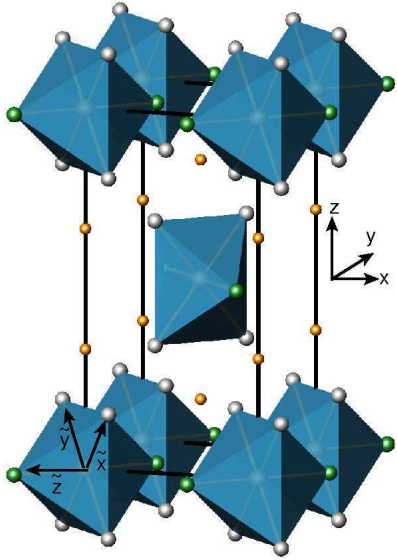}\end{center}
\caption{\label{structure}(Color Online) The tetragonal trirutile
structure of CuSb$_{2}$O$_{6}$. There are
two Cu atoms per formula unit (two sublattices). 
The CuO$_{6}$ octahedra in these 2 sublattices are rotated by
90$^{\degree}$ with respect to each other. Both the global 
($x$,$y$,$z$) and the local ($\tilde{x},\tilde{y},\tilde{z}$)
coordinate systems used in our calculations are depicted. 
Within the local coordinate system,
the in-plaquette oxygens are highlighted in grey while the out-of-plaquette
oxygens are in green. The Sb atoms are highligted in yellow. The in-plaquette Cu-O bondlength is 0.04 \AA larger than
the out-of-plaquette Cu-O bondlength.    }
\end{figure}

The tetragonal structure\cite{giere} ($a$ = 4.629 \AA\ , $c$ = 9.288 \AA) 
of CuSb$_{2}$O$_{6}$ with two formula units per unit cell
is elucidated in Fig. \ref{structure}.
One interesting aspect of this system is the bond lengths, the
out-of-plaquette Cu-O bond length is shorter by 0.04 \AA\ compared to
the in-plaquette Cu-O bond lengths. The CuO$_{6}$ octahedra are well
separated from each other, with no edge-sharing or corner-sharing
between adjacent CuO$_{6}$ octahedra.  An angle $\beta \approx$
91.2$\degree$ is the main distinguishing factor of the monoclinic (T $<$
380K ) phase of CuSb$_{2}$O$_{6}$.  Nevertheless, the calculated
density of states (DOS) and band structure for both these phases are
nearly identical. Therefore, all band structure calculations shown here have been
performed using the tetragonal tri-rutile structure instead of the
monoclinic one for the sake of simplicity.  Using the more regular tetragonal structure is 
quite advantageous in calculating and understanding the tight-binding
hopping parameters and in the construction of larger supercells to
model specific magnetic orderings.

We performed band structure calculations within the local (spin)
density approximation (L(S)DA) using the full-potential local orbital
minimum basis code FPLO 
(version 5.00-18).\cite{fplo1,footnote1}
In the
scalar relativistic calculations the exchange correlation potential of
Perdew and Wang\cite{PW92} was used. Additionally, the strong Coulomb
repulsion in the Cu 3$d$ orbitals was taken into account in a mean
field approximation by LSDA+$U$ calculations\cite{czyzyk}, using the
around-mean-field double-counting scheme. 
Variations of $U_{d}$ ($J$ = 1eV) within the relevant
physical range ($U_{d}$ = 5 to 9 eV) did not change the reported results significantly.

\begin{figure}[t]
\begin{center}\includegraphics[%
  clip,
  width=8.5cm,
  angle=0]{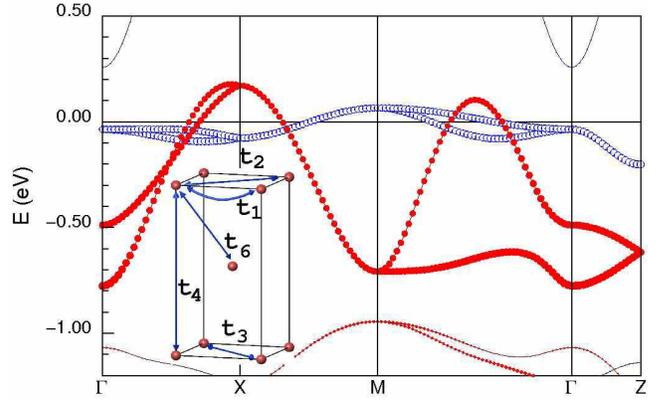}\end{center}
\caption{\label{bands}(Color Online) LDA band structure of the
tetragonal CuSb$_{2}$O$_{6}$ along high-symmetry directions. All four
$e_{g}$ bands are partially filled. The `narrow' band (open circle,
blue) is entirely from the local Cu $d_{\tilde{x}^{2}-\tilde{y}^{2}}$
orbital, while the `broad' band (filled circle, red) comes from the Cu
$d_{3\tilde{z}^{2}-\tilde{r}^{2}}$ orbital. The inset shows the leading
hopping terms of the tight-binding model (see text).}
\end{figure}

The calculated non-magnetic LDA band structure, displayed in
Fig.~\ref{bands}, shows four bands crossing the Fermi level
E$_{F}$, i.e. two pairs of narrow and broad bands.
Considering the two formula units per cell, this is a very unusual result. 
In the standard cuprate scenario, only one well separated half-filled antibonding 
band per Cu$^{2+}$ is expected.
Projecting the orbital character onto the band structure in a local
coordinate system wherein the local $\tilde{z}$-axis is perpendicular to the
Cu-O plaquette-plain (see Fig.~\ref{structure}) and points towards the
apical oxygens, we find that the `narrow' bands originate from the
in-plaquette Cu $d_{\tilde{x}^{2}-\tilde{y}^{2}}$ orbitals, while the `broad' bands come
from the out-of-plaquette Cu $d_{3\tilde{z}^{2}-\tilde{r}^{2}}$ orbital.  This is 
different from the conventional cuprate scenario where only one band
per Cu atom crosses the Fermi level and is comprised predominantly of
the in-plaquette Cu $3d$ orbital character.  This unique feature stems
from the fact that the CuO$_{6}$ octahedra are only slightly distorted
in  CuSb$_2$O$_6$, so that the cubic degeneracy for the $e_g$
ligand-field states are only slightly lifted. The energy difference between the
$d_{\tilde{x}^{2}-\tilde{y}^{2}}$ and $d_{3\tilde{z}^{2}-\tilde{r}^{2}}$ 
related band centers is about 0.3 eV
only, to be compared with about 2 eV  for `standard' cuprates.

Because of the strong competition between the in- and out-of-plaquette orbitals,
the effects of strong correlations (LSDA+$U$) on the magnetic model 
are not immediately obvious. 
Let us first review the possible ways in which  1D
order can be invoked in this system. {\bf Scenario 1:} The interaction of the electrons 
is mediated via the CuO$_{4}$ plaquette, 1D chains will be formed along the 
[1$\bar{1}$0] (and along [110] in the other sublattice).
{\bf Scenario 2:} The interaction of the electrons is mediated via the 
out-of-plaquette oxygen. This would in turn create 1D chains along the [110]
(along [1$\bar{1}$0] in the other sublattice) direction.
These two competing scenarios in principle could lead to 
the same spin ordering pattern of 1D antiferromagnetic chains.
Therefore strong correlations may drive the system to either 
one of these two scenarios.

Firstly, to compare the microscopic magnetic interactions in CuSb$_{2}$O$_{6}$,
we did a two-site one-band TBM fit, 
separately for the d$_{\tilde{x}^{2}-\tilde{y}^{2}}$ and
d$_{3\tilde{z}^{2}-\tilde{r}^{2}}$ band complexes. The respective 2x2 tight-binding matrix is given by,
\[
\left( \begin{array}{cc}
 \alpha + \beta & \delta \\
 \delta & \alpha + \beta^{\prime} \\
\end{array} \right)
\]
with
\begin{eqnarray*}
\alpha &=& \epsilon_{0} + 2t_{1}\left[C_{\tilde{x}} + C_{\tilde{y}}\right] + 
           2t_{4}C_{\tilde{z}} + 4t_{5}\left[C_{\tilde{x}} + C_{\tilde{y}}\right]C_{\tilde{z}} \\
\beta  &=& 2t_{2}\left[C_{\tilde{x}}C_{\tilde{y}} - S_{\tilde{x}}S_{\tilde{y}}\right] + 
           2t_{3}\left[C_{\tilde{x}}C_{\tilde{y}} + S_{\tilde{x}}S_{\tilde{y}}\right] \\
\beta^{\prime} &=& 2t_{3}\left[C_{\tilde{x}}C_{\tilde{y}} - S_{\tilde{x}}S_{\tilde{y}}\right] + 
           2t_{2}\left[C_{\tilde{x}}C_{\tilde{y}} + S_{\tilde{x}}S_{\tilde{y}}\right] \\
\delta &=& 8t_{6}C_{\frac{1}{2}\tilde{x}}C_{\frac{1}{2}\tilde{y}}C_{\frac{1}{2}\tilde{z}} + 8t_{7}C_{\frac{1}{2}\tilde{z}}\left[C_{\frac{3}{2}\tilde{x}}C_{\frac{1}{2}\tilde{y}} - C_{\frac{1}{2}\tilde{x}}C_{\frac{3}{2}\tilde{y}}\right]\\
\end{eqnarray*} 
where $C_{\tilde{x}}$ = $\cos(k_{\tilde{x}}a)$,
$C_{\frac{1}{2}\tilde{x}}$ = $\cos(k_{\tilde{x}}\frac{a}{2})$,
$C_{\frac{3}{2}\tilde{x}}$ = $\cos(k_{\tilde{x}}\frac{3a}{2})$,
$S_{\tilde{x}}$ = $\sin(k_{\tilde{x}}a)$  and the
corresponding cyclic permutations. The values of the transfer
integrals obtained from the fit are collected in
Table.~\ref{hoppings}, with the hopping paths shown in the inset of
Fig.~\ref{bands}.  The hopping parameters obtained from the fit to the
narrow d$_{\tilde{x}^{2}-\tilde{y}^{2}}$ band complex are all quite
similar in magnitude, implying almost equal interaction strength along
the three main hopping paths t$_{1}$, t$_{3}$, and t$_{4}$ (see
Table.~\ref{hoppings}. This is indicative of a 3D magnetic model and
the energy scale of the transfer integrals are too small. This 3D
model is incompatible with the experimental data.\cite{gibson} The fit
to the broad d$_{3\tilde{z}^{2}-\tilde{r}^{2}}$ band produces a
transfer integral (t$_{2}$) which is very large compared to all others
(see Table.~\ref{hoppings}).  This suggests a strong 1D AFM
interaction along the [110] direction, implying that the superexchange
path Cu-O-O-Cu is along the [110] basal diagonal and is via the apical
oxygens and NOT via the in-plaquette oxygens.  This is in line with
the band characters of the projection mentioned above.  This leads to
almost perfect 1D chains of antiferromagnetic Cu atoms along the [110]
direction. Another interesting fact to notice is the difference in the
next-nearest-neighbor interaction along [110] ($t_{2}$) and
[$\bar{1}$10] ($t_{3}$). The Cu-O bondlength along [110] is shorter (2
\AA\ ) with a 180$^{\degree}$ bond angle as compared to the Cu-O
bondlength of 3.5 \AA\ along [$\bar{1}$10] and a 160$^{\degree}$
Cu-O-Cu bond angle.  This structural feature also helps in the one
dimensionality along [110].  The individual exchange constants
calculated using J$_{ij}$ = 4t$^{2}_{ij}$/U$_{eff}$, with a U$_{eff}$
= 4.5 eV\cite{footnote2} are also collected in Table.~\ref{hoppings}.
The sublattice coupling, connecting the two Cu sites in the unit cell
is rather small, but not negligible, and so are the interchain
couplings. This is very much consistent with the 1D nature of the
system.  Though the plaquettes are quite isolated in
CuSb$_{2}$O$_{6}$, they can interact in many different ways and this
in turn determines the magnitude of the exchange constants. But the
unique geometry in this system gives a very large 1D exchange
(J$_{2}^{TBM}$ = 400K) as compared to other 2D cuprates. This value of
J$_{2}$ is too large (by a factor of four) due to possible
ferromagnetic contributions like in other cuprate compounds. We
perform LSDA+$U$ calculations to get a deeper insight.

Within LSDA+$U$, depending on the choice of the input
density matrix, we were able to correlate (fill the spin-up band and unfill
the spin-down band) both band complexes in the vicinity of the Fermi level. Contrary to 
other typical cuprates, the ground state is obtained when correlating the 
`broad' Cu $d_{3\tilde{z}^{2}-\tilde{r}^{2}}$ orbital (i.e.
$d_{\tilde{x}^{2}-\tilde{y}^{2}}^{\uparrow}$, $d_{\tilde{x}^{2}-\tilde{y}^{2}}^{\downarrow}$, 
$d_{3\tilde{z}^{2}-\tilde{r}^{2}}^{\uparrow}$ :
filled and  $d_{3\tilde{z}^{2}-\tilde{r}^{2}}^{\downarrow}$ : empty) 
with an energy gain of about 
110 meV ($\sim$ 1280 K)\cite{footnote3}
per Cu atom 
as compared to the `narrow' $d_{\tilde{x}^{2}-\tilde{y}^{2}}$ orbital (i.e.
$d_{3\tilde{z}^{2}-\tilde{r}^{2}}^{\uparrow}$, $d_{3\tilde{z}^{2}-\tilde{r}^{2}}^{\downarrow}$, 
$d_{\tilde{x}^{2}-\tilde{y}^{2}}^{\uparrow}$ :
filled and  $d_{x^{2}-y^{2}}^{\downarrow}$ : empty). 
This result is in accordance with the second scenario described above and also
with the 1D TBM result. 
Superexchange
via the apical Cu-O orbitals is quite surprising and unique, though not 
categorically excluded. Since the apical Cu-O bondlength is slightly shorter than
the in-plaquette one, the covalency of Cu with the apical oxygens are slightly
larger than that with the planar oxygens.  We can also notice from the LDA band structure 
(Fig.~\ref{bands}) that the $d_{\tilde{x}^{2}-\tilde{y}^{2}}$ band is quite narrow compared to
 $d_{3\tilde{z}^{2}-\tilde{r}^{2}}$ band. Keeping this in mind, we can interpret the unique
orbitally ordered ground state to be the result of the relative gain in 
kinetic energy when correlating the broad Cu $d_{3\tilde{z}^{2}-\tilde{r}^{2}}$ orbital.   

\begin{figure}[t]
\begin{center}\includegraphics[%
  clip,
  width=9.0cm,
  angle=0]{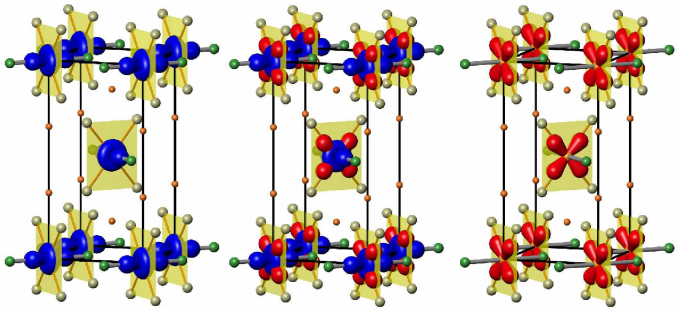}\end{center}
\caption{\label{orbitals}(Color Online) {\bf Middle panel:}
The competition of in-plaquette and out-of-plaquette orbitals in LDA for CuSb$_{2}$O$_{6}$. Both orbitals
are partially occupied and comprise the Fermi surface.
{\bf Right panel:} Standard picture of high T$_c$ cuprates with the inclusion of strong Coulomb correlations $U$.
The plaquette orbitals remain half-filled and thereby would lead to 3D magnetic order.
{\bf Left panel:} Unique orbital ordering in CuSb$_2$O$_6$ leading to a strong 1D magnetic order. }
\end{figure}

Our TBM for the Cu $d_{3\tilde{z}^{2}-\tilde{r}^{2}}$ orbitals results in  
very short-ranged interactions (Table.~\ref{hoppings}). Therefore an 
extension of the supercells beyond the second neighbors is not necessary.
LSDA+$U$ calculations of differently ordered spin configurations (FM and AFM)
were performed to obtain
an effective exchange constant, J$_{2}^{nn}$ by mapping
the energies obtained from the calculations to a 
Heisenberg model (neglecting the other J's since they are small
according to the TBM),
$H = \sum_{\langle i,j \rangle } J_{ij} \mathbf{S}_i \cdot \mathbf{S}_j$ which leads to  
$E_{fm} - E_{afm} = 2J_{2}^{nn}|S|^2 , S=\frac{1}{2}$.
This model leads to  J$_{2}^{nn}$ $\sim$ 140 K, in good comparison to the value 
of about 100K obtained in susceptibility experiments.\cite{gibson}
The slight overestimation of the leading exchange is rather typical for this
type of calculation\cite{johannes} and also somewhat dependent on the choice of $U$ 
which is not exactly known. We obtain a charge transfer gap of 2.2 eV which is
consistent with the green color of the sample. 
\begin{center}
\begin{table}
\begin{tabular*}{0.47\textwidth}%
     {@{\extracolsep{\fill}}|c|c|c|c|c|c|c|c|}
\hline
(meV) & t$_1$ & t$_2$ & t$_3$ & t$_4$ & t$_5$ & t$_6$ & t$_7$\tabularnewline
\hline
 d$_{x^{2}-y^{2}}$ & -20 & - & 17.5 & 20.8 & 3.80 & - & -1.86 \tabularnewline
\hline
d$_{3z^{2}-r^{2}}$ & 9.52 & -197 & -13.1 & -3.86 & - & -17.9 & - \tabularnewline
\hline
\end{tabular*}
\vspace{0.5cm}
\begin{tabular*}{0.47\textwidth}%
     {@{\extracolsep{\fill}}|c|c|c|c|c|c|c|c|}
\hline
 (K) & J$_1$ & J$_2$ & J$_3$ & J$_4$ & J$_5$ & J$_6$ & J$_7$ \tabularnewline
\hline
d$_{x^{2}-y^{2}}$ & 4.23 & - & 3.15 &  4.46 & 0.15 & - & 0.04 \tabularnewline
\hline
d$_{3z^{2}-r^{2}}$ & 0.93 & 400 & 1.76 & 0.15 & - & 3.32 & - \tabularnewline
\hline
\end{tabular*}
\caption{\label{hoppings}Hopping parameters and the corresponding 
exchange constants from a two-site one-band 
TBM. The hopping paths used are: t$_1$ [000]$\rightarrow$[100],
 t$_2$ [000]$\rightarrow$[110], t$_3$ [000]$\rightarrow$[$\bar{1}$10], 
t$_4$ [000]$\rightarrow$[001], t$_5$ [000]$\rightarrow$[101],
t$_6$ [000]$\rightarrow$[$\frac{1}{2}\frac{1}{2}\frac{1}{2}$], 
t$_7$ [000]$\rightarrow$[$\frac{3}{2}\frac{1}{2}\frac{1}{2}$].
The hopping paths are indicated in the inset of Fig.~\ref{bands}. 
The fit has been done for both the in-plaquette
d$_{\tilde{x}^{2}-\tilde{y}^{2}}$ and the out-of-plaquette d$_{3\tilde{z}^{2}-\tilde{r}^{2}}$ bands 
separately. The parameters are vanishingly small when no value is
provided. }
\end{table}
\end{center}
The presence of quite regular octahedra in CuSb$_{2}$O$_{6}$ introduces
an orbital degree of freedom among the $e_{g}$ orbitals as compared to other
standard cuprates.
Electronic structure calculations give convincing evidence for the
strong competition between the d$_{\tilde{x}^{2}-\tilde{y}^{2}}$ and d$_{3\tilde{z}^{2}-\tilde{r}^{2}}$  orbitals
for the ground state in CuSb$_{2}$O$_{6}$.  Correlations
drive a unique d$_{3\tilde{z}^{2}-\tilde{r}^{2}}$ orbital ordering pattern in the 
ground state (Fig.~\ref{orbitals}). Our TBM  and total energy calculations 
(mapped onto an effective Heisenberg hamiltonian) are in favor of
a 1D (d$_{3z^{2}-r^{2}}$) instead of 3D (d$_{x^{2}-y^{2}}$) scenario following 
a particular orbital ordering. Our ordering pattern is consistent with the
magnetic unit cell proposed from the results of neutron diffraction experiments.\cite{nakua,gibson}
This in turn elucidates that the 
low-lying excitations for cuprates are not restricted {\it a priori} to the standard
orbital picture.  

The analysis above has only considered the isotropic superexchange interaction. 
Allowing for spin-orbit coupling introduces anisotropic terms in the superexchange interaction.
Many model calculations have been performed for various cuprates to determine the
magnetic anisotropy energy,\cite{hayn99} which scales inversely with the difference
between the d$_{x^{2}-y^{2}}$ and d$_{3z^{2}-r^{2}}$ energy levels. Contrary to other
cuprates, this energy difference in CuSb$_{2}$O$_{6}$ is extremely small due to the presence
of a rather regular octahedron. Therefore, we can qualitatively infer that CuSb$_{2}$O$_{6}$ exhibits a large 
magnetic anisotropy.
This anisotropy reduces the phase space of the isotropic Heisenberg model and should lead to 
a significant increase of the magnetic ordering temperature. Indeed, if 
we estimate the N{\'e}el temperature (T$_{N}$) assuming an isotropic
$S = \frac{1}{2}$ quasi-one-dimensional chain within the bosonization 
method\cite{irkhin,footnote4},
\[
T_{N} = 0.33kJ^{\prime}z_{\perp}\sqrt{\ln\frac{5.8J}{T_{N}} + \frac{1}{2}\ln\ln\frac{5.8J}{T_{N}}} 
\]
we find T$_{N}^{theo}$ $\approx$ 1.5 K which strongly underestimates
the observed T$_{N}^{expt}$ $\approx$ 8.5 K.  Due to the mean
field-like treatment of the inter-chain coupling, T$_{N}^{theo}$ is
expected to be an upper estimate.\cite{rosner97} The strong
underestimation of the ordering temperature using the isotropic model
implies the presence of considerable anisotropy in the system, which
is naturally explained by our orbital scenario.
  
In conclusion, electronic structure and TBM calculations in CuSb$_{2}$O$_{6}$ reveal 1D
behavior from orbital ordering due to a competition between the Cu $d_{\tilde{x}^{2}-\tilde{y}^{2}}$ and
$d_{3\tilde{z}^{2}-\tilde{r}^{2}}$ orbitals; strong Coulomb repulsion selects the latter (Fig.~\ref{orbitals}).
The main reason for such a ground state is the underlying local environment. Our results
are in agreement with the susceptibility and recent neutron diffraction measurements for the spin structure.\cite{gibson}
Performing polarization dependent x-ray absorption spectroscopy (XAS) should
elucidate the choice of orbital ordering in CuSb$_{2}$O$_{6}$. A strong 
anisotropy in the absorption edge will be observed when using the incoming
radiation E $\parallel$ {\bf a-b} plane as compared to E $\parallel$ {\bf c}. 
We are currently analyzing the results from this experiment. 
Effects of pressure on this system are quite interesting, as they can flip the balance between the two competing 
orbitals thereby restoring the standard plaquette picture as in other cuprates.

We acknowledge critical remarks from
A. Ormeci, W. Ku, M.~D. Johannes. Funding from the Emmy-Noether
program is acknowledged.




\begin{references}
\bibitem{footnote}All the experimental exchange constants have been normalized to sum over
bonds to aid easy comparison to our results.
\bibitem{footnote1} The extension of valence orbitals (Cu $3d4s4p$, Sb $5s5p5d$ and O $2s2p3d$) were
optimized with respect to total energy using an additional confining
potential ~ $(r/r_{0})^{4}$.
\bibitem{footnote2}For the isolated CuO$_{6}$ geometry we use a representative $U_{eff}$ = 4.5 eV according to reference 
\onlinecite{johannes}.
\bibitem{footnote3}Comparing this energy gain (1276 K) for 
correlating the Cu $d_{3\tilde{z}^{2}-\tilde{r}^{2}}$ orbital with that of
J$_{2}^{TBM}$ = 400 K, allows for the separation of the spin and orbital 
degrees of freedom due to varied energy scales.
\bibitem{footnote4}$z_{\perp}$ = 4 is the number of neighbors in transverse to chain 
directions and $k$ for a tetragonal lattice (dimension $d$ = 1+2) is set to 
0.7.\cite{irkhin} 
The variable $J$ denotes the leading coupling along the direction of the
chain, set equal to $J_{expt}$ = 100K, and $J^{\prime}$ denotes 
non-frustrating interchain coupling, set equal to $\sqrt{J_{3}J_{4}}$ = 0.5 K, using
the results from our TBM following the procedure of Ref. \onlinecite{rosner97}.
\end{references}
\end{document}